\documentclass[nofootinbib,twocolumn,aps]{revtex4}

\usepackage{amsmath}
\usepackage{amssymb}

\DeclareMathOperator*{\argmax}{argmax}

\usepackage{graphicx}
\usepackage{dcolumn}
\usepackage{bm}
\usepackage{hyperref}

\setcitestyle{numbers,square}

\begin{document}

\title{A General Bayesian Algorithm for the Autonomous Alignment of Beamlines}

\author{T.~W.~Morris} 
 \email{tmorris@bnl.gov}
 \altaffiliation[also at ]{LBNL}
\author{M.~Rakitin}
\author{Y.~Du}
\author{M.~Fedurin}
\author{A.~C.~Giles}
\author{D.~Leshchev}
\author{W.~H.~Li}
\author{B.~Romasky}
 \altaffiliation[also at ]{Stony Brook University}
\author{E.~Stavitski}
\author{A.~L.~Walter}
\affiliation{Brookhaven National Laboratory, Upton, NY 11973}

\author{P.~Moeller}
\author{B.~Nash}
\affiliation{RadiaSoft LLC, Boulder, CO 80301}

\author{A.~Islegen-Wojdyla}
\affiliation{Lawrence Berkeley National Laboratory, Berkeley, CA 94720}

\date{\today}

\begin{abstract}

Autonomous methods to align beamlines can decrease the amount of time spent on diagnostics, and also uncover better global optima leading to better beam quality. The alignment of these beamlines is a high-dimensional, expensive-to-sample optimization problem involving the simultaneous treatment of many optical elements with correlated and nonlinear dynamics. Bayesian optimization is a strategy of efficient global optimization that has proved successful in similar regimes in a wide variety of beamline alignment applications, though it has typically been implemented for particular beamlines and optimization tasks. In this paper, we present a basic formulation of Bayesian inference and Gaussian process models as they relate to multiobjective Bayesian optimization, as well as the practical challenges presented by beamline alignment. We show that the same general implementation of Bayesian optimization with special consideration for beamline alignment can quickly learn the dynamics of particular beamlines in an online fashion through hyperparameter fitting with no prior information. We present the implementation of a concise software framework for beamline alignment and test it on four different optimization problems for experiments at x-ray beamlines of the National Synchrotron Light Source II and the Advanced Light Source and an electron beam at the Accelerator Test Facility, along with benchmarking on a simulated digital twin. We discuss new applications of the framework, and the potential for a unified approach to beamline alignment at synchrotron facilities.

\end{abstract}


\maketitle

\section{Introduction}


Synchrotron light sources are invaluable scientific tools that allow the probing of materials across bulk, micron, and nanometer scales. 
These facilities perform a wide variety of research, with applications in the study of catalysis, biological function, and material science. 
Several next-generation synchrotron and free-electron laser facilities are slated for upgrades which will increase their brilliance by several orders of magnitude \citep{borland2018upgrade, chenevier2018esrf, galayda2018lcls, white2019new}.
%
However, more advanced experiments will require more precise and complex optical setups. 
%
Beamlines consist of a large number of optical components (e.g. mirrors, magnets, apertures), each with many degrees of freedom (corresponding to e.g. motors that translate, rotate, and bend the components).\footnote{See e.g. Figure~\ref{fig:tes-schematic}.} These degrees of freedom can be highly correlated or degenerate, making beamline alignment in essence a high-dimensional ($D~\gtrsim~10$), highly non-linear optimization problem.
%
This is typically done manually,\footnote{The design of optical systems is typically done to separate some of these dimensions and make manual alignment more feasible, e.g. by prefocusing and refocusing with a secondary-source aperture and a pair of Kirkpatrick-Baez mirrors.} but as the complexity and precision of beamlines grow, the development of efficient and robust automated alignment methods is necessary for the efficient operation of light sources now and in the future. 
Such methods allow us to reach an acceptable level of alignment more quickly and robustly than with manual methods when realignment is necessary, saving preparation and commissioning time which could be used for experiments.
%
They further allow us to potentially find better global optima than an operator could discover manually by considering all dimensions of the beamline simultaneously.
%
They also represent the first step toward a fully autonomous beamline \citep{maffettone2023artificial}.

Some attempts at beamline alignment apply methods like genetic and differential evolution \citep{xi2015general, xi2017ai, rakitin2020introduction, zhang2023multi}, attempt to match beamline data to an online model \citep{nash2022combining, nash2022online},
or use families of commonly-used optimization algorithms \citep{breckling2022automated, morris2022fly}. These approaches are limited in that they give no guarantee of convergence to a global optimum. They also make no consideration of minimizing the number of function evaluations, and beamline optimization almost universally involves a prohibitively expensive-to-sample function, both at the real beamline (relying on the movement of precise motors, which can be slow) and on simulated digital twins (relying on computationally intensive ray-tracing \citep{sanchez2011shadow3} or Fourier-based methods \citep{chubar2013wavefront}) meaning that their use is intractable for large numbers of dimensions.
In contrast to the classical methods above, algorithms based on machine learning construct and fit a model to understand the effects of changing the parameter inputs, as well as the interaction of the output beam qualities (e.g. flux, spatial resolution, energy resolution, polarization, coherence), leading to a more efficient search of the parameter space.
Some machine learning methods like reinforcement learning \citep{velotti2022automatic} suffer from similar drawbacks to the methods above in that they may take too long to learn enough to be useful, at which point beamline parameters and hyperparameters may have drifted substantially.
From a practical point of view, then, we should greatly prefer alignment methods that converge as quickly as possible, and rely on little to no prior input.


%

A machine learning framework well-suited for expensive-to-sample functions is Bayesian optimization, which performs well with no prior information on optimization problems that are expensive-to-sample, high-dimensional, and potentially very noisy. Bayesian optimization has been applied in such a wide variety of contexts as synchrotron light sources \citep{rebuffi2023autofocus, morris2023latent}, free-electron lasers \citep{duris2020bayesian}, particle colliders \citep{cisbani2020ai}, and laser–plasma-based ion sources \citep{dolier2022multi}. These implementations, however, are typically applicable to single experiments; indeed, much of the difficulty in implementing machine learning solutions to any problem is the trade-off of specificity and generality where an algorithm that is specific enough to be effective in some context is too specific to be applied generally.


Bayesian optimization is highly generalizable in the choice of the kernel model used to describe the parameter space, and that many facilities are moving toward shared software environments and shared data acquisition protocols like Bluesky \citep{allan2019bluesky, rakitin2022next} suggest the benefit of a general agent. This paper demonstrates an implementation of a Bayesian agent that can learn the dynamics and idiosyncrasies of a particular beamline and can thus be deployed across many different beamlines with relatively little implementation cost by applying the same code to a range of optimization problems at different synchrotron and non-synchrotron facilities.

In Sections \ref{sec:bo}, \ref{sec:gp}, and \ref{sec:acquisition} we present a general but brief formulation of multiobjective Bayesian optimization with Gaussian process regression as it relates to this work (for a more thorough introduction see \citealt{frazier2018tutorial}). Section~\ref{sec:beamlines} addresses beamline-specific considerations for Bayesian optimization, and Section~\ref{sec:code} presents their implementation in a software package. Section~\ref{sec:experiments} describes the application of the code at beamlines across the National Synchrotron Light Source II and Accelerator Test Facility at Brookhaven National Laboratory and the Advanced Light Source at Lawrence Berkeley National Laboratory, and presents the results of benchmarking on a simulated digital twin. Finally, Section~\ref{sec:discussion} discusses the future development of the algorithm.

\section{Bayesian optimization}
\label{sec:bo}

\begin{figure*}
\centering
\includegraphics[width=.99\textwidth]{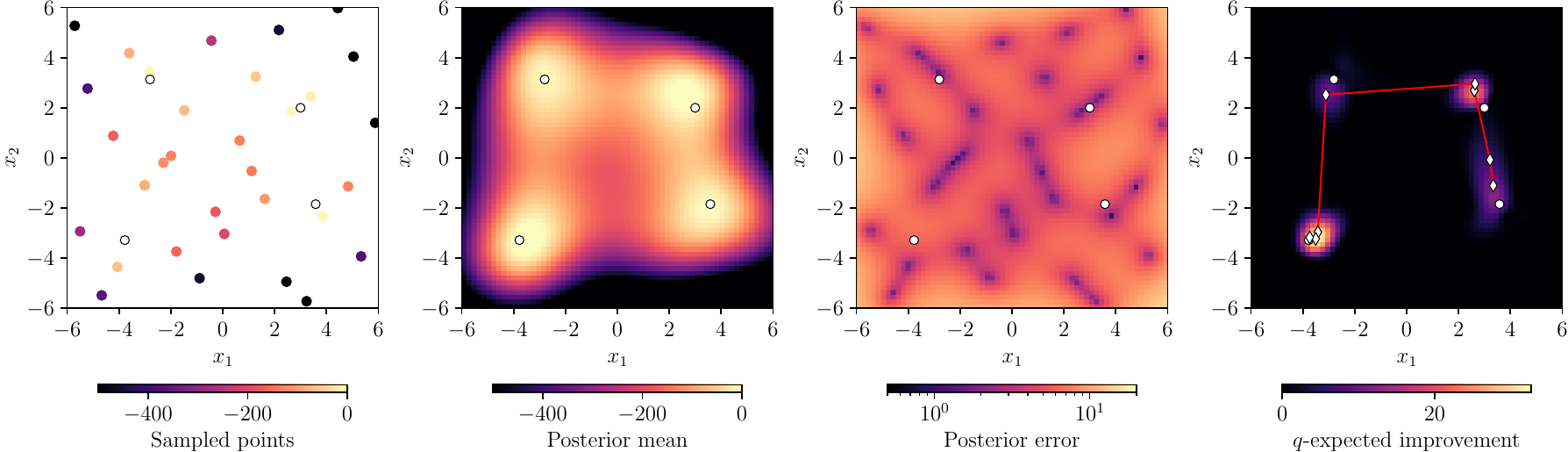}%
\caption{An example of an iteration of a Bayesian optimization algorithm trying to maximize the negated Himmelblau's function $f(x_1,x_2)= - (x_1^{2}+x_2-11)^{2} - (x_1+x_2^{2}-7)^{2}$, whose true global optima are marked as white circles. Using existing data points (\textit{far left}) and the assumption that the function is distributed as a Gaussian process, we can use Bayesian inference to compute a posterior consisting of a mean (\textit{center left}) and error (\textit{center right}), upon which we can compute an acquisition function (\textit{far right}) which informs us of the best points to sample. The black-edged diamonds superimposed on the acquisition function show the best eight points to sample, optimized in parallel, and the optimal routing represented by the red line.}
\label{fig:himmelblau}
\end{figure*}

\begin{figure*}
\centering
\includegraphics[width=.99\textwidth]{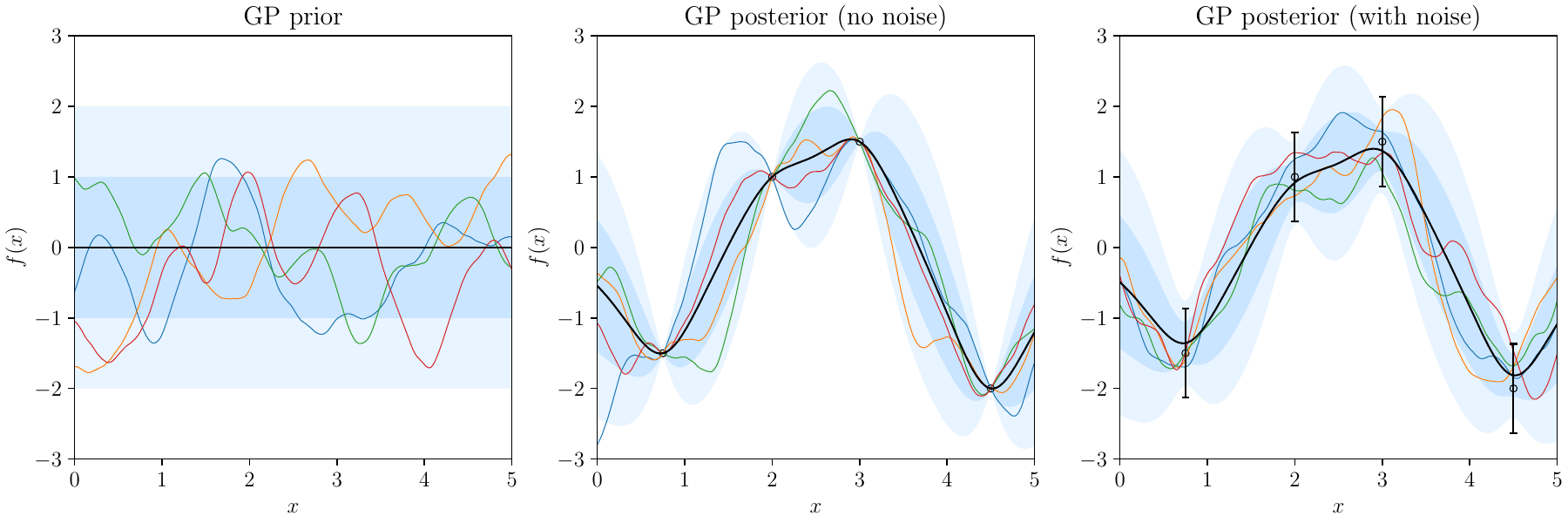}
\caption{The prior distribution, noiseless posterior distribution, and noisy posterior distributions for a Gaussian process with covariance $\langle f(x_i) f(x_j) \rangle = M_{5/2}(|x_i - x_j|/4)$, where $M_\nu(r)$ is the Matérn function as defined in Equation~\ref{eqn:matern}. For each distribution, we draw four random functions (colored lines). The black line represents the mean of each distribution, while the dark- and light-shaded regions represent the 1$\sigma$ and 2$\sigma$ intervals.}

\label{fig:gp-prior-and-posterior}
\end{figure*}

Consider an expensive-to-sample black-box function $f(x)$ with $d$-dimensional inputs $x \in \mathbb{R}^d$. In finding the right input $x$ to achieve the maximal value of $f(x)$, it is untenable to utilize optimization methods that rely on lots of function samples. We can address this by treating the function as a stochastic process (which describes a distribution over all possible realizations of the function) and using Bayesian inference to construct a posterior distribution $p(f)$, i.e. describing how likely it is that every possible function $f$ is the true function.\footnote{In the case of no noise, the support of $p(f \mid x, y)$ consists of only those functions $f$ for which $f(x) = y$.} If we sample the function at points $x = \{x_1, x_2,...,x_n\}$ and observe values $y = \{f(x_1), f(x_2),...,f(x_n)\}$, then we can use Bayesian inference to write our posterior belief about $f$ given that we observe $x$ and $y$ as
\begin{equation}
\label{eqn:bayesian_inference}
    p(f \mid x, y) = \frac{p(y \mid f, x) p(f)}{p(y \mid x)}
\end{equation}
where the quantity $p(y \mid f, x)$ (called the \textit{likelihood}) is the probability of observing values $y$ at inputs $x$ for a given function $f$, the quantity $p(f)$ (called the \textit{prior}) is our knowledge about the probability of a given function $f$ before we have seen any data, and the quantity $p(y \mid x)$ (called the \textit{marginal likelihood}), represents the distribution of $y$ after marginalizing over the distribution of $f$.\footnote{We know that the marginal likelihood must be
\begin{equation}
\label{eqn:marginal_likelihood}
    p \big ( y \mid x \big ) = \int p(y \mid f, x) p(f) df
\end{equation}
because the sum of all the probabilities in the posterior must sum to unity.} A representation of conditioning a prior on observations to construct a posterior is shown in Figure~\ref{fig:gp-prior-and-posterior}. Each iteration of Bayesian optimization then consists of three steps:
\begin{enumerate}
    \item Estimate the posterior $p(f \mid y, x)$ from some historical observations $(x, y)$.
    \item Use the posterior to find the most desirable point $x^\star$ within some predefined bounds.
    \item Sample that point, and add it to our historical observations. 
\end{enumerate}
Constructing a posterior from observations in the first step is almost always done with a Gaussian process (GP), the particulars of which are described in Section~\ref{sec:gp}. Quantifying the desirability of candidate points in the second step is done using acquisition functions which are described in Section~\ref{sec:acquisition}. A concrete example of an iteration of Bayesian optimization as applied to minimizing Himmelblau's function is shown in Figure~\ref{fig:himmelblau}, using a GP model and an acquisition function that computes the expected improvement in the cumulative maximum by sampling each candidate point.


\section{Gaussian process models}
\label{sec:gp}




A Gaussian process (GP) is a stochastic process where every collection of variables $y$ has a multivariate normal distribution; for notational simplicity and without loss of generality, we assume throughout the paper that all of our processes are zero-mean. The GP is described entirely by the covariance matrix $\Sigma$ describing the observations $y$. A Gaussian process model consists of assigning a covariance matrix to a set of sample data $y$ at inputs $x$ and computing the posterior mean and posterior variance at every other input. In practice, the covariance of the process is not known \textit{a priori} and is approximated by constructing and fitting a kernel.

\subsection{Kernels and hyperparameter optimization}

We model the covariance matrix with a kernel matrix $K(x, x', \theta)$, where
\begin{equation}
    K_{ij} = k(x_i, x_j, \theta)
\end{equation}
where $k$ is a kernel function, $x_i$ and $x_j$ are two inputs, and $\theta$ is a set of hyperparameters which tune $k$. The only constraint on a kernel matrix $K$ is that it is positive-definite.\footnote{A positive-definite matrix is a symmetric matrix whose eigenvalues are all strictly positive.} A simplifying assumption is to require that the kernel is stationary, that is, that the correlation of the function at two inputs depends only on their distance
\begin{equation}
k(x_i, x_j, \theta) = k(|x_i - x_j|, \theta)
\end{equation}
To construct our kernel, we take the hyperparameters which maximize the marginal likelihood
\begin{equation}
    \hat{\theta} = \argmax_\theta p \big ( f(x) \mid \theta \big )
\end{equation}
For a Gaussian process, the marginal likelihood is given by
\begin{multline}
    p \big ( y \mid x, \theta \big ) = \exp \Big [ - \frac{1}{2} \big ( y^\dagger K(x, x, \theta)^{-1} y \\- \log \det K(x, x, \theta) + n \log 2 \pi \big ) \Big ]
\end{multline}
\begin{equation}
\label{eqn:multivariate-gaussian-pdf}
    p(y) = \frac{1}{\sqrt{(2\pi)^n | \Sigma|}} \exp { \big ( - \frac{1}{2} y^\dagger K(x, x, \theta)^{-1} y \big ) }
\end{equation}

\subsection{Posterior estimation}

Once we have our kernel $K(x_i, x_j, \theta)$ and optimized hyperparameters $\hat{\theta}$, we can use Gaussian process regression to construct posteriors. Given our measurements $y$ at points $x$, our posterior estimate of the distribution of the process at points $x^\star$ is a Gaussian distribution with posterior mean $\big \langle f(x^\star) \big \rangle = A f(x)$ and posterior covariance $\big \langle f(x^\star) \otimes f(x^\star) \big \rangle = B$, where
\begin{equation}
\label{eqn:A}
A = K(x^\star, x, \hat{\theta}) K(x, x, \hat{\theta})^{-1}
\end{equation}
\begin{equation}
\label{eqn:B}
B = K(x^\star, x^\star, \hat{\theta}) - A K(x, x^\star, \hat{\theta})
\end{equation}
The vector of variances for each individual point is the diagonal of the posterior covariance $B$.

\subsection{Noisy models}

It may be the case that our observations are noisy, i.e. that observing the function at points $x$ will yield $y = f(x) + \epsilon$ where $\epsilon$ is a random noise term. If we assume that $\epsilon$ is homoskedastic and Gaussian, then we can account for the noise by adding a constant noise variance $\sigma^2$ to the diagonal of the kernel $K$.\footnote{A small noise level (or ``jitter") is desirable even for noiseless GP models to make the Cholesky decomposition of the kernel a well-conditioned problem.}

\section{Acquisition functions}
\label{sec:acquisition}

The acquisition function $\mathcal{A}(x)$ is a model of a given objective over possible inputs which, given a posterior $p(f \mid x, y)$, quantifies the desirability of sampling a given input $x$. For each iteration of the optimization, we optimize the acquisition function over the inputs as
\begin{equation}
    x^* = \argmax_x \mathcal{A}\big ( p(f \mid x, y) \big )
\end{equation}
Acquisition functions over posteriors are typically cheap to compute, and so classical algorithms (like LM-BFGS) are used to optimize them.\footnote{In regimes of lots of data, however, computing and optimizing acquisition functions in parallel can be computationally expensive. We note the benefits of GPU-accelerated acquisition function optimization, though we do not implement it in this work.}Acquisition functions can be either analytic or non-analytic; below, we show benefits and examples of either approach.

\subsection{Analytic acquisition functions}

Analytic acquisition functions are directly computable from the posterior; as the posterior for a GP is determined entirely by the mean $\mu$ and variance $\sigma$, they may be expressed as
\begin{equation}
    \mathcal{A}(x) = f \big ( \mu(x), \sigma(x) \big )
\end{equation}
The simplest example is the expected mean
\begin{equation}
    \text{EM}(x) = \mu(x),
\end{equation}
where on every iteration the algorithm will sample the point with the largest expected mean. A less risk-averse example is the expected improvement 
\begin{equation}
    \text{EI}(x) = \Big \langle \max \big ( f(x) - f^\star, 0 \big ) \Big \rangle 
\end{equation}
which is our expectation for how much the cumulative maximum $f^\star$ will increase if we were to sample $x$. We can compute this directly as 
\begin{equation}
    \text{EI}(x) = \int_{f^*}^{\infty} y \phi(y) dy = \sigma(x) \big ( \phi(z) + z \Phi(z))
\end{equation}
where $z = (\mu(x) - f^*) / \sigma(x)$, and $\phi(z)$ and $\Phi(z)$ are the PDF and CDF of the standard normal distribution. Because repeated sampling of a point will strictly decrease the posterior variance, this algorithm will (for well-behaved problems) eventually explore every point in the parameter space.

\subsection{Monte Carlo acquisition functions}

Some useful acquisition functions cannot be computed directly from the mean and variance of the posterior. Acquisition functions that involve sampling from the posterior to estimate some ensemble are more flexible and often more robust. One example of this is in selecting multiple points, as in when we want to find the best $n$ points to sample given some analytic acquisition function $\mathcal{A}(x)$: presumably they should be spread out to better cover the parameter space, but there no obvious way to quantify and thus compute that analytically. We address this interdependence with a Monte Carlo acquisition function, where we might evaluate the acquisition of some collection of points by sampling from the posterior and taking an ensemble average of the result. There is a large benefit in sampling multiple points at once for beamline optimization, as it allows us to find a batch of points to sample and then optimally route the beamline parameters between them to reduce travel time.

Note that all analytic acquisition functions have a Monte-Carlo equivalent; an example is shown in the far-right panel of Figure~\ref{fig:himmelblau}, where we use the $q$-expected improvement as an acquisition function to find a parallel set of eight inputs to sample next.\footnote{The $q$ refers to the $q$-batching formalism used to denote the axis of Monte Carlo samples.} Monte Carlo methods also allow for more sophisticated information theory-based acquisition functions like Predictive Entropy Search \citep{hernandez2014predictive}, Maximum Entropy Search \citep{wang2017max}, or Joint Entropy Search \citep{hvarfner2022joint}.

\subsection{Multiobjective optimization}
\label{sec:multiobjective}

Optimization problems often require managing trade-offs. For example, a common beamline design consists of a secondary-source aperture (SSA), which cuts off some flux in the interest of having a smaller and tighter beam. One method of multiobjective optimization is scalarization, which is to use a function that maps a vector output to a scalar output, leading to one quantity to be maximized. In this work, we use affine scalarizations (i.e. assigning a weight to each objective and summing them) to construct a single fitness function over which to optimize an acquisition function. 

There are, however, are other useful ways to carry out multiobjective optimization like Pareto efficient searches (i.e. finding the set of inputs where no one objective can be increased without decreasing some other objective). But while fully multiobjective methods do allow for more flexibility in the alignment, that flexibility is not compatible with an autonomous beamline, which must decide on a single best beam and thus must collapse the beam to a single fitness function.

\section{Beamline-specific considerations}
\label{sec:beamlines}

In this section, we consider beamline-specific considerations that improve the practical application of Bayesian optimization to the automated alignment problem. In this paper, we consider the common optimization problem of maximizing the beam power density, defined as 
\begin{equation}
    \rho(x) = \Phi(x) / (\sigma_x(x) \sigma_y(x))
\end{equation}
where $x$ represents the beamline inputs to optimize and where $\Phi(x)$, $\sigma_x(x)$, and $\sigma_y(x)$ are the input-dependent flux, horizontal spread, and vertical spread of the beam. These parameters are inferred from an image of the beam profile, taken using either an area detector (e.g. Figures \ref{fig:iss-beams} and \ref{fig:als-beams}) or a beam stop and microscope (e.g. Figures \ref{fig:tes-beams} and \ref{fig:atf-beams}). 
In practice, it is better to model the fitness as
\begin{equation}
    \log \rho(x) = \log \Phi(x) - \log \sigma_x(x) - \log \sigma_y(x)
\end{equation}
because the distribution in variations in beam flux and size are both roughly log-normal, and so their logarithms are better described by a Gaussian process. It also preserves the convexity of the problem and, being inherently dimensionless, allows us to affinely scalarize many simultaneous objectives as a single Gaussian process.

\subsection{A kernel for latent beamline dimensions}
\label{sec:latent}

\begin{figure}
\centering
\includegraphics[width=.5\textwidth]{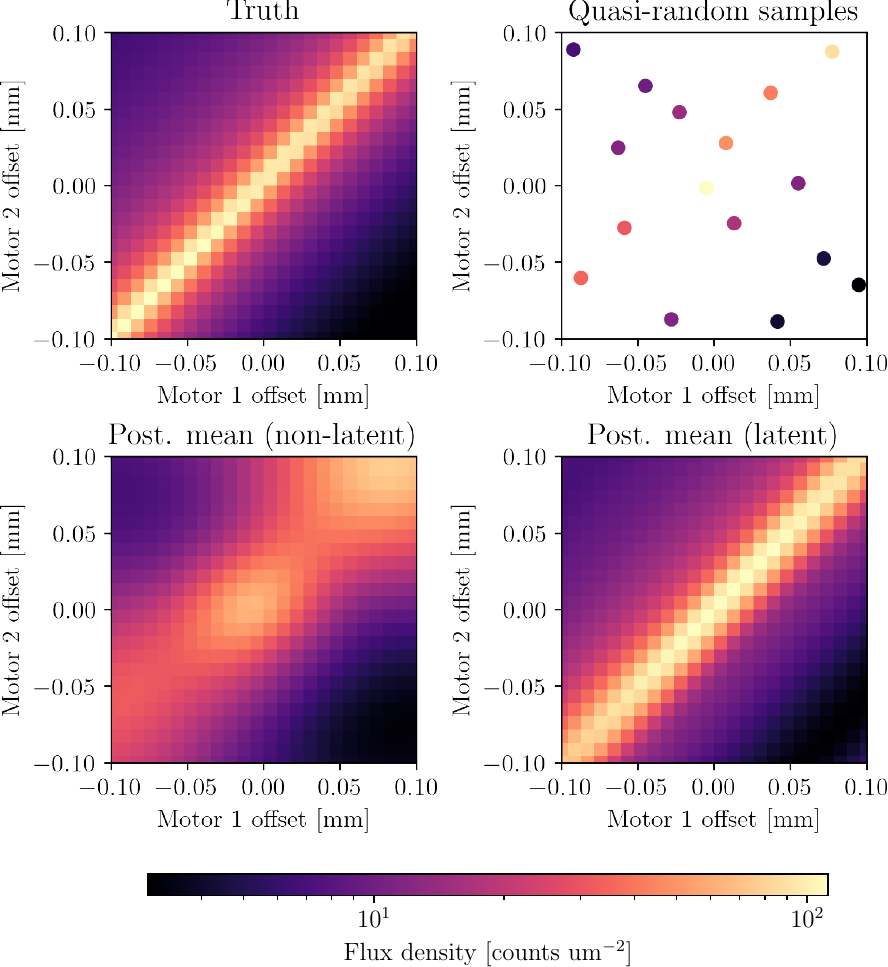}
\caption{\textit{Upper left:} the result of changing the positions of two coupled dimensions of the TES beamline. \textit{Upper right:} a quasi-random sample of 16 points from the ground truth. \textit{Lower left}: a non-latent GP fitted to the parameter space fitted to the sampled points. \textit{Lower right}: a latent GP fitted to the same points, which correctly infers the latent dimensions.}
\label{fig:coupled_dimensions}
\end{figure}

Input parameters for beamlines can be highly coupled, as shown in Figure~\ref{fig:coupled_dimensions}. In fitting Gaussian processes to beamline data, we adopt a kernel of the form 
\begin{equation}
\label{eqn:kernel_design}
    k(x_i, x_j, \theta) = f \Big ( \big | D \exp S  \big ( x_i - x_j \big ) \big | \Big )
\end{equation}
where $f(r)$ is some radial function, $D$ is a diagonal matrix with positive entries, $\exp(\cdot)$ is the matrix exponential, and $S$ is a skew-symmetric matrix. Because the matrix exponential of a real skew-symmetric matrix is an orthogonal matrix, this kernel represents a norm-preserving transformation in the parameter space by $\exp(S)$ and a scaling of each dimension in the new basis by $D$. The hyperparameters $\theta$ define the entries of $D$ and $S$, which for an $d$-dimensional parameter space have $d$ and $d(d - 1)/2$ degrees of freedom respectively, together defining a total transformation matrix $T = D \exp (S)$ with $d(d + 1)/2$ degrees of freedom. 

This kernel design is guaranteed to be positive-definite so long as $f$ is a positive-definite function.\footnote{Per Bochner's theorem, a function $f$ is a positive-definite function if it is the Fourier transform of a weakly positive function on the real line.} A commonly used positive-definite function in kernel construction is the Matérn function, which can be written as
\begin{equation}
    f(r) = a^2 \big ( r / \ell \big )^\nu K_\nu \big ( r / \ell \big ) 
\end{equation}
where $K_\nu(z)$ is the modified Bessel function of the second kind of order $\nu$, and where $a, \ell, \nu > 0$ are hyperparameters. For our purposes, $\ell$ as a lengthscale parameter is redundant and can be subsumed into the hyperparameters defining $T$. This leaves us with a normalized form
\begin{equation}
\label{eqn:matern}
    f(r) = M_{\nu}(r) = a^2 r^\nu K_\nu \big ( r \big ) 
\end{equation}
Bessel functions are expensive to compute for arbitrary $\nu$, so we constrain our kernel to $\nu = n + 1/2, n \in \mathbb{Z}$, for which Equation~\ref{eqn:matern} reduces to the product of a polynomial and an exponential. Unless otherwise specified, we use $\nu = 5/2$ throughout the paper. 

\subsection{Dirichlet-based validity constraints}
\label{sec:constraints}

The application of Bayesian optimization relies on reliable diagnostic feedback, which is often not a realistic assumption for real-life scenarios. Undesirable behavior in the diagnostics can occur both sporadically (e.g. in the case of a beam dump or a hardware failure) and also systematically (a certain beamline orientation causes the beam to miss a mirror or detector). We want to be able to classify regions of the parameter space as invalid and encode that knowledge into our acquisition function, but we don't want a single unrepresentative glitch to rule out an otherwise worthy part of the parameter space. For this purpose a probabilistic classification model is ideal, and it is also able to easily adjust expectation- and entropy-based acquisition functions. 

We use the classification method outlined in \cite{milios2018dirichlet} which fits a Dirichlet distribution to the data from which we can generate class probabilities. This method has the benefit of avoiding expensive posterior sampling (as in the case with stochastic variational methods) at the expense of not being able to quantify uncertainty. 

A Dirichlet distribution of order $N$ is defined as
\begin{equation}
    f(p, \alpha) = \frac{\Gamma \big ( \sum_i \alpha_i \big )}{\prod_i \Gamma(\alpha_i)} \prod_i p_i^{\alpha_i-1}
\end{equation}
where the vector parameter $p = \{p_1, ..., p_N\}$ describes the probabilities of classes in a $N-1$ simplex (so that $\sum_i p_i = 1$), and the concentration parameters $\alpha = \{\alpha_1, ..., \alpha_N\}$ parameterize the concentration of the distribution in that simplex. We use transformed Gaussian processes to model the concentration parameter $\alpha_i$ for each classification $i$ according to \cite{milios2018dirichlet}. As the order-$N$ Dirichlet distribution is the conjugate prior to the $N$-categorical distribution, we can obtain the probability of a beamline input being valid as 
\begin{equation}
    \pi_i(x) = \frac{\gamma_i}{\sum \gamma_i}, \quad \gamma_i \sim \text{Gamma}(\alpha_i, 1).
\end{equation}
where $\text{Gamma}(\alpha, \beta)$ is the Gamma distribution, and $\gamma_i$ are samples from that distribution. Using this probability, we can weight any objective-based acquisition function to prefer inputs that lead to valid outputs. 
This approach has the added benefit of being generalizable to any number of classification labels, which could be made more nuanced than a binary model of validity.

\subsection{Sampling expense}

Bayesian optimization is particularly useful when sampling the objective function $f(x)$ is expensive.
%
This is strictly true for some beamlines where computing a diagnostic is expensive, e.g. those that involve intensive data processing or a complicated metaroutine like a knife-edge scan \citep{ji2019knife}.
%
Many beamlines, though, have no latency in the diagnostics and are only expensive to sample because they are expensive to move around.
%
This is due to the high precision of the motors, which must move slowly so as not to damage the optics, and need time to settle to prevent backlash and make sure that it is exactly at its setpoint; this is typically on the order of several seconds.

A good acquisition function, then, should take into account travel time: a simple solution is to optimize a Monte Carlo acquisition function over a ``batch" of points that we can compute the most efficient route between using e.g. \url{https://github.com/google/or-tools}. Ideally, the acquisition function would consider the variable time cost of traveling to a given set of points, but the computational cost of this can be unwieldy. 
%

\subsection{Hysteresis}
\label{sec:hysteresis}

Another challenge to ML-based optimization is hysteresis, which manifests at beamlines when the actual position of some input varies from the desired input. This can happen when the motor approaches the same position from different directions, primarily from physical backlashes in the hardware. 
A core assumption of Bayesian optimization is that the relevant function $f(x)$ always yields the same output (modulo some noise). 
Hysteresis can be mitigated by overestimating the noise level, or with a more thorough treatment of uncertainty in the inputs of the underlying Gaussian process \citep{liu2022robust}.
We note the benefit of motor encoders, which can lead to more precise and consistent control of beamline hardware.

\subsection{Composite objectives}

Even though we combine estimates of the different beam attributes into a scalar fitness to maximize, it is still beneficial to construct and train three separate models for the flux, horizontal spread, and vertical spread, a method typically referred to as \textit{composite optimization}. This allows us to take advantage of how different inputs affect different outputs; indeed, many beamlines are designed to separate components that tune the flux from those that tune the focus. This can significantly reduce the effective dimensionality of the alignment problem.


\section{Implementation}
\label{sec:code}

\subsection{Beamline Optimization (Blop)}
\label{sec:blop}

Our beamline alignment tools are implemented in the Blop\footnote{See \url{https://nsls-ii.github.io/blop} for documentation and tutorials.} Python package, relying on the BoTorch Python package \citep{balandat2020botorch}.
In Blop we develop a customized kernel which fits to latent beamline dimensions outlined in Section~\ref{sec:latent}, and weight common acquisition functions by the probabilistic constraint outlined in Section~\ref{sec:constraints}. 
We also use BoTorch for model fitting and acquisition function optimization.
The algorithm is used in terms of an agent, which we instantiate with motors and diagnostic equipment. We can ``tell" the agent about the values of pre-defined objectives (e.g. beam height, coherence) and ``ask" it for new points to sample. 
The agent wraps the steps of Bayesian optimization in a single customizable routine (implemented in Python as a {\tt .learn()} method), which yields a plan accepted by the Bluesky experiment orchestration system.\footnote{In effect, a single button that can be pressed to align the beamline.} This routine can be tailored to each beamline (or to each alignment problem for a given beamline). Encapsulating the optimization as a single process simplifies the alignment from the point of view of the user, making experimentation more accessible to users who have less familiarity with a given beamline (i.e. hardware, data acquisition, control systems), or with less familiarity with software in general.

\subsection{Bluesky}

Bluesky \citep{allan2019bluesky, rakitin2022next} is a software package that allows for the orchestration and execution of experiments from Python, and is in the process of being adapted by various light sources.\footnote{See \url{https://blueskyproject.io/}} We design Blop with Bluesky in mind, as it can use Bluesky to automatically take data, analyze it, and optimize the inputs with the same feedback and control systems used for beamline experiments. This allows the Blop agent to both command and control the beamline, leading to an easier implementation.\footnote{Though Blop is not limited to Bluesky facilities and can be made to only ``command" the experiments using only its ``ask" and ``tell" methods.} Bluesky has been mainly developed by NSLS-II, with a growing international collaboration at multiple facilities where it is used and expanded. The adoption of a single standard for experimental control and analysis across many facilities allows us to apply the same automated alignment tools with relatively little effort.

\section{Experiments}
\label{sec:experiments}

\subsection{Alignment of a Kirkpatrick-Baez mirror system at the TES beamline}
\label{sec:real-tes}

\begin{figure}[h!]
\centering
\includegraphics[width=.5\textwidth]{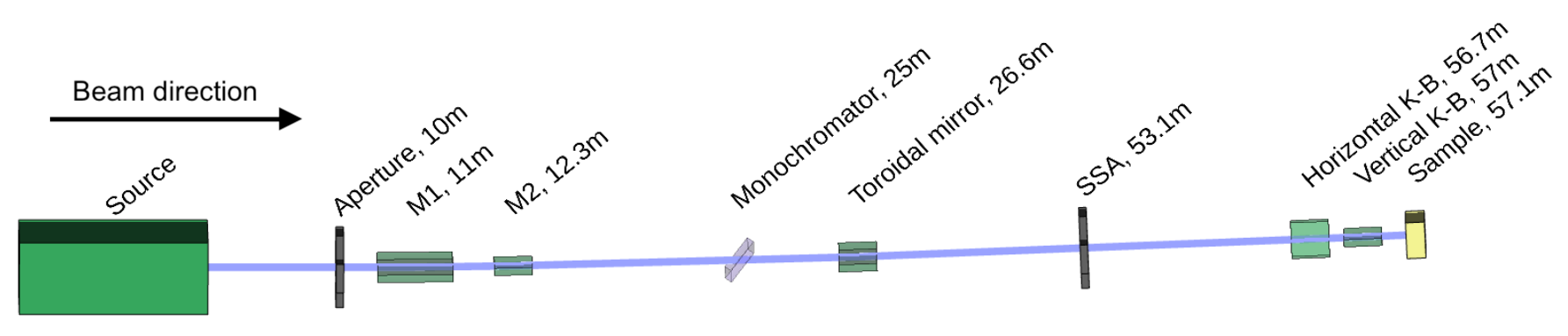}%
\caption{A schematic of the Tender-Energy X-ray Absorption Spectroscopy (TES, 8-BM) beamline at NSLS-II. This representation shows the many optical components that make up modern beamlines, with each optical component having many degrees of freedom that must be optimized in concert in order to carry out experiments effectively.}
\label{fig:tes-schematic}
\end{figure}

\begin{figure}
\centering
\includegraphics[width=.5\textwidth]{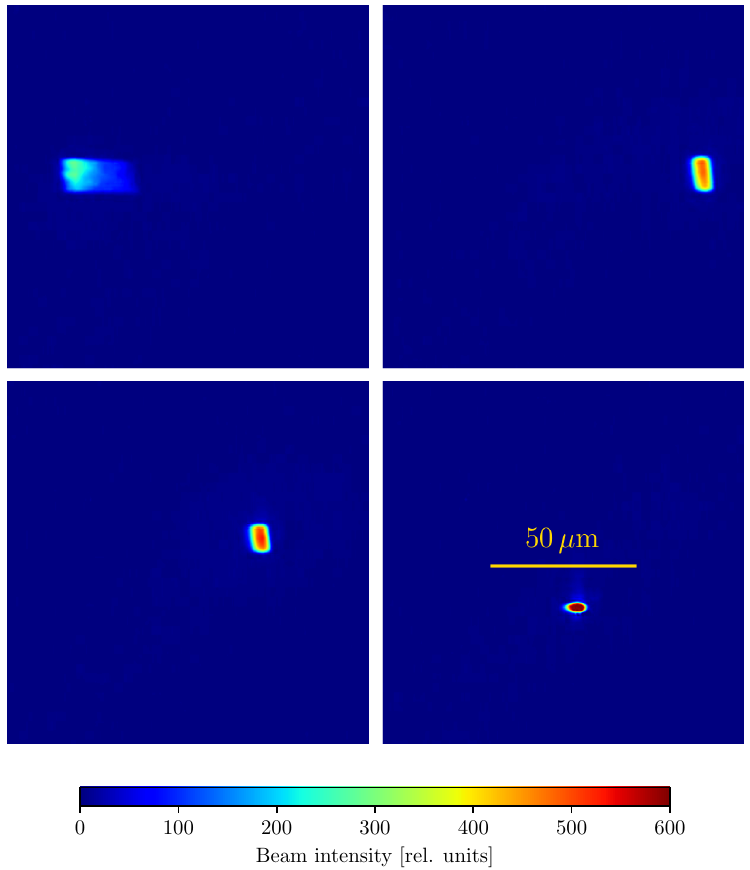}
\caption{Four different beams configurations at the National Synchrotron Light Source II's Tender Energy X-ray Absorption Spectroscopy (TES, 8-BM) beamline, where the upper left panel shows the initial beam and where the lower right represents the optimal alignment. In this alignment test, we adjust the translation and rotation of each of the horizontal and vertical Kirkpatrick-Baez mirrors and the pitch and vertical translation of a toroidal mirror, for a total of six degrees of freedom to maximize the flux density of the beam.}
\label{fig:tes-beams}
\end{figure}

The TES beamline \citep{northrup2019tes} is a tender x-ray microspectroscopy beamline at NSLS-II with an energy range of 2-5.5 keV and a beam size which can be tuned between 5 and 20 microns. The x-rays are produced from a bending magnet source and pass through a Si(111) double crystal monochromator. A toroidal mirror prefocuses the beam onto a secondary source aperture (SSA), after which the beam is refocused onto the sample by a pair of Kirkpatrick-Baez (K-B) mirrors in the endstation chamber. A schematic of the beamline is represented in Figure~\ref{fig:tes-schematic}.
We optimize for the flux density on the sample by allowing each K-B mirror and the toroidal mirror to pitch and translate into and out of the beam for a total of six degrees of freedom. Figure~\ref{fig:tes-beams} shows an example of the beam feedback provided by the camera, with the alignment being gradually improved. 


\subsection{Alignment of a Johann Spectrometer at the ISS beamline}

\begin{figure}
\centering
\includegraphics[width=.5\textwidth]{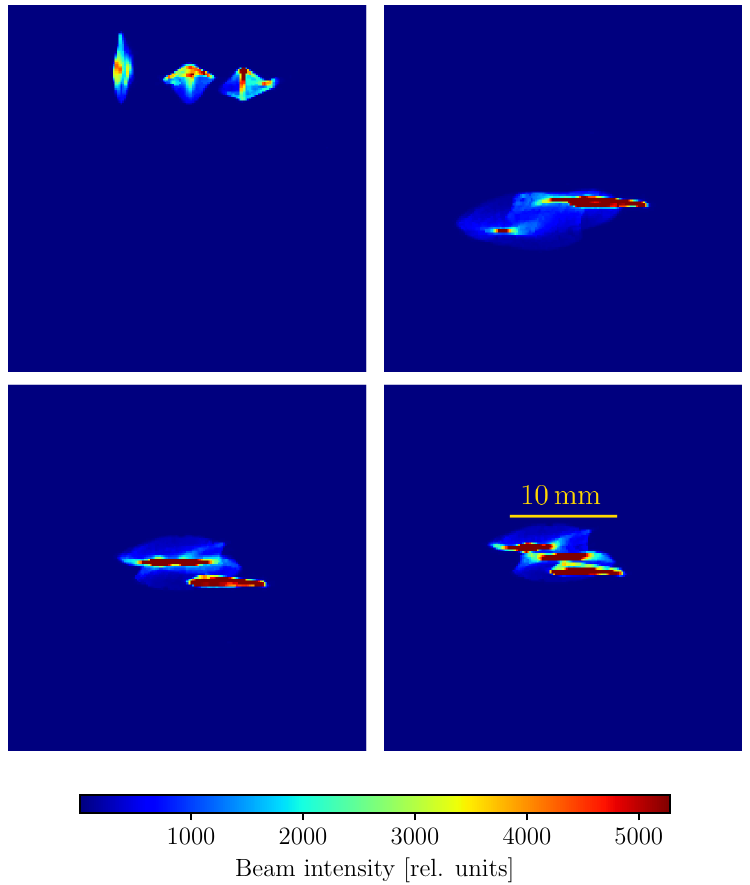}
\caption{Four different beam configurations at the National Synchrotron Light Source II's Inner Shell Spectroscopy (ISS, 8-ID) beamline during automated alignment, where the upper left panel shows the initial beam and the lower right panel represents the optimal alignment. In this alignment test, we adjust the translation of a central crystal and the translation and pitch of two ancillary crystals for a total of five degrees of freedom to maximize the flux density of the total beam on the area detector.}
\label{fig:iss-beams}
\end{figure}

The Inner Shell Spectroscopy beamline (ISS, 8-ID) beamline \citep{leshchev2022inner} is a beamline for x-ray absorption spectroscopy and operando and in-situ characterization of materials. ISS is a damping wiggler beamline with a Si(111) monochromator capable of producing energies between 4.9 and 33~keV. Currently, the beamline is developing high-resolution capabilities with the recent commissioning of a five-analyzer Johann-type spectrometer, where after hitting the sample, the beam is reflected back onto an area detector by several crystals.\footnote{For an overview and schematic of Johann-type spectrometers, see \cite{kleymenov2011five}.} Maximizing the flux on the area detector maximizes the resolution of the spectrometer, and so we seek to colocate the reflections of the crystals onto the same point. We use three crystals to focus the beam onto a two-dimensional area detector. Figure~\ref{fig:iss-beams} shows the optimization of the three-crystal system.





\subsection{Photon transport optimization at the Advanced Light Source beamline 5.3.1}

Beamline 5.3.1 at the Advanced Light Source at Lawrence Berkeley National Laboratory is a research and development beamline. It is a bending magnet beamline, operating in the tender x-ray regime (2.4–12 keV photon energy range.), where the instrument controls have recently been upgraded to the EPICS/Bluesky framework.

\begin{figure}[h!]
\centering
\includegraphics[width=.5\textwidth]{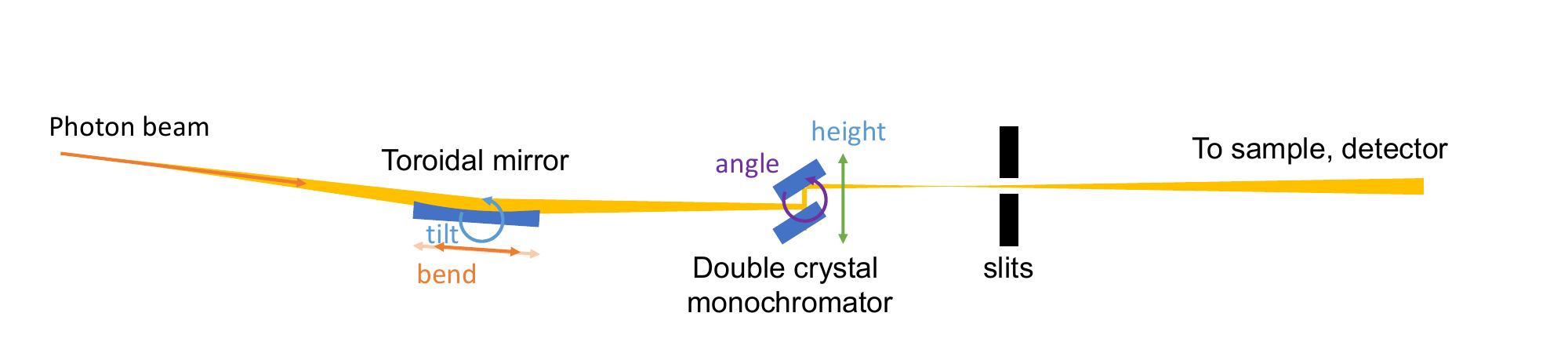}
\caption{A schematic of beamline 5.3.1 at the Advanced Light Source. The beamline has four degrees of freedom (toroidal mirror pitch and bend; monochromator angle and height), and four constraints (4-jaw slits).}
\label{fig:bl531-schematic}
\end{figure}

The photon transport system (Figure~\ref{fig:bl531-schematic}) comprises of a first focusing mirror, a monochromator and a few apertures. The focusing mirror is a vertically deflecting toroidal mirror, creating an image of the source at the sample. The mirror is gold-coated with a nominal grazing angle of 5~mrad, mirror-to-object ($p$) and mirror-to-image ($q$) distance of $p=q=12$\,m. The corresponding tangential and sagittal radius of curvature are respectively $R_t = 2400$\,m and $R_s = 60$\,mm. The mirror is bendable along the tangential direction to adjust the vertical focus position. The monochromator is a channel-cut double crystal Si(111) monochromator, providing a 25~mm vertical offset. There are a set of 4-jaw slits immediately after the monochromator to block the straight-through beam, and another set of 4-jaw slits immediately before the sample position (12\,m downstream of the toroidal mirror).

\begin{figure}[h!]
\centering
\includegraphics[width=.5\textwidth]{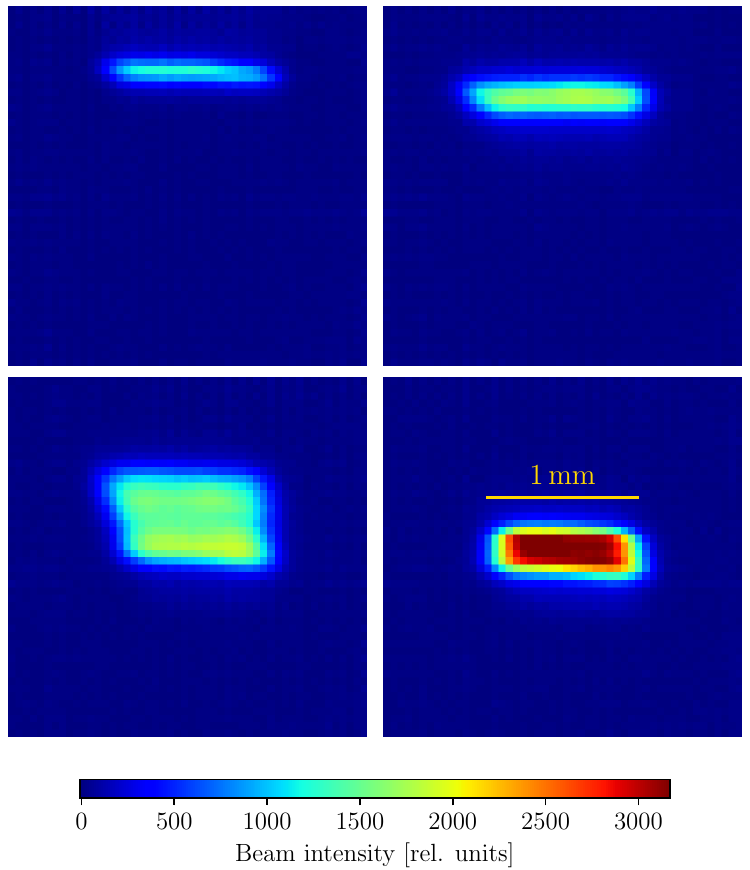}
\caption{Four different beam configurations at the Advanced Light Source's Beamline 5.3.1 during automated alignment. In total, the photon transport has four active degrees of freedom: the focusing mirror pitch and tangential bend, the channel cut crystal angle and height. The upper left panel shows the initial, manually aligned beam and the lower right the final beam after automated alignment. Upper right and lower left panels show intermediary points collected in the automated alignment process.}
\label{fig:als-beams}
\end{figure}

For beam measurement, we used a diamond-based x-ray beam monitor (ClearXCam from Advent Diamond) with which we computed the flux as the sum of all pixels. 
We added a preference for a rounder beam (with some coupling between horizontal and vertical size) by defining  an ``effective area" metric as:
\begin{equation}
\label{eqn:effective_area}
    \text{EA}(x) = \sigma^2_\text{width}(x) + \sigma^2_\text{height}(x)
\end{equation}
The full scalarized fitness for the effective power density then becomes
\begin{equation}
\label{eqn:alt-fitness}
    f(x) = \log f(x) - \log \text{EA}(x).
\end{equation}
The manual optimization is rendered difficult by the interplay between the toroidal mirror angle, monochromator height and angle, all of them changing the beam height and interfering with 4-jaw slits. With the use of the described automated alignment, we were able to maximize the power density on the sample in under five minutes, with a final beam size of 1~mm x 0.3~mm (HxV, FWHM), close to the theoretical limit calculated by raytracing  (Figure \ref{fig:als-beams}) with an improvement of the intensity by over a factor of two over our best effort using manual alignment.

\subsection{Alignment of an electron beam at the Accelerator Test Facility}

\begin{figure}
\centering
\includegraphics[width=.5\textwidth]{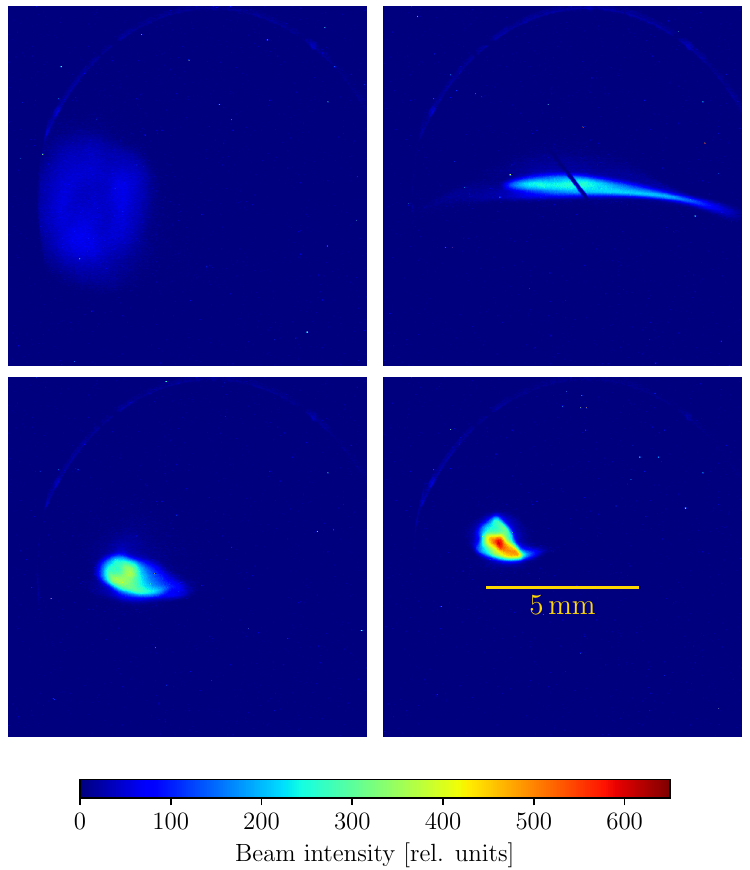}
\caption{Four different electron beam configurations at the Brookhaven National Laboratory's Accelerator Test Facility (ATF) at different stages of automated alignment, where the upper left panel shows the starting beam and the lower right the optimal beam. In this alignment test, we tune the current of four quadrupole electromagnets to maximize the objective in Equation~\ref{eqn:alt-fitness}.}
\label{fig:atf-beams}
\end{figure}

The Accelerator Test Facility (ATF) is a user facility at Brookhaven National Laboratory offering the combination of an 80~MeV electron beam synchronized with a Terawatt picosecond carbon dioxide laser \citep{pogorelsky2014brookhaven}. This gives it the capability to develop cutting-edge electron beam techniques including ultrafast electron diffraction and microscopy \citep{mcdonald1988design}, as well as free-electron laser techniques including direct laser acceleration and using Compton scattering as a high energy x-ray source \citep{batchelor1990microwiggler}. We modulate three bending quadrupole electromagnets and a solenoid to manipulate the shape of the beam, for a total of four degrees of freedom. 

We employ the alternate fitness function in Equation~\ref{eqn:alt-fitness} that was also used to align ATF. Figure~\ref{fig:atf-beams} shows an example of the beam feedback provided by the in-house beam diagnostic, with the alignment being gradually improved. 

\subsection{Simulated alignment of the TES beamline}

\begin{figure}
\centering
\includegraphics[width=.5\textwidth]{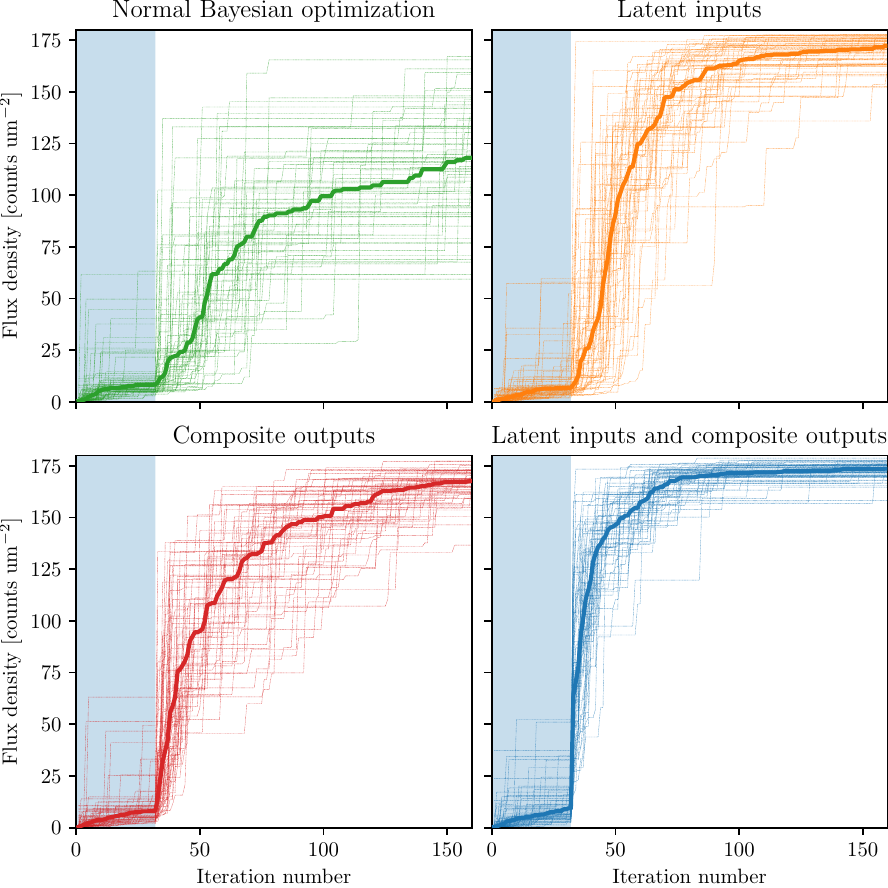}
\caption{The 8-dimensional optimization of the simulated TES beamline, where the degrees of freedom comprise the toroidal and Kirkpatrick-Baez mirrors. The colors show different varieties of Bayesian optimization algorithms with and without both of latent inputs and composite outputs, with both the cumulative maximum of all individual runs (thin lines) and the median cumulative maximum (thick line). Each variety starts out with a quasi-random sampling of 32 points (shaded light blue), and then performs a Bayesian optimization loop with the expected improvement acquisition function. The benefit of using both latent inputs and composite outputs is shown, as we can achieve a better optimum more robustly and more quickly. 
}
\label{fig:tes-convergence-8d}
\end{figure}

\begin{figure}
\centering
\includegraphics[width=.5\textwidth]{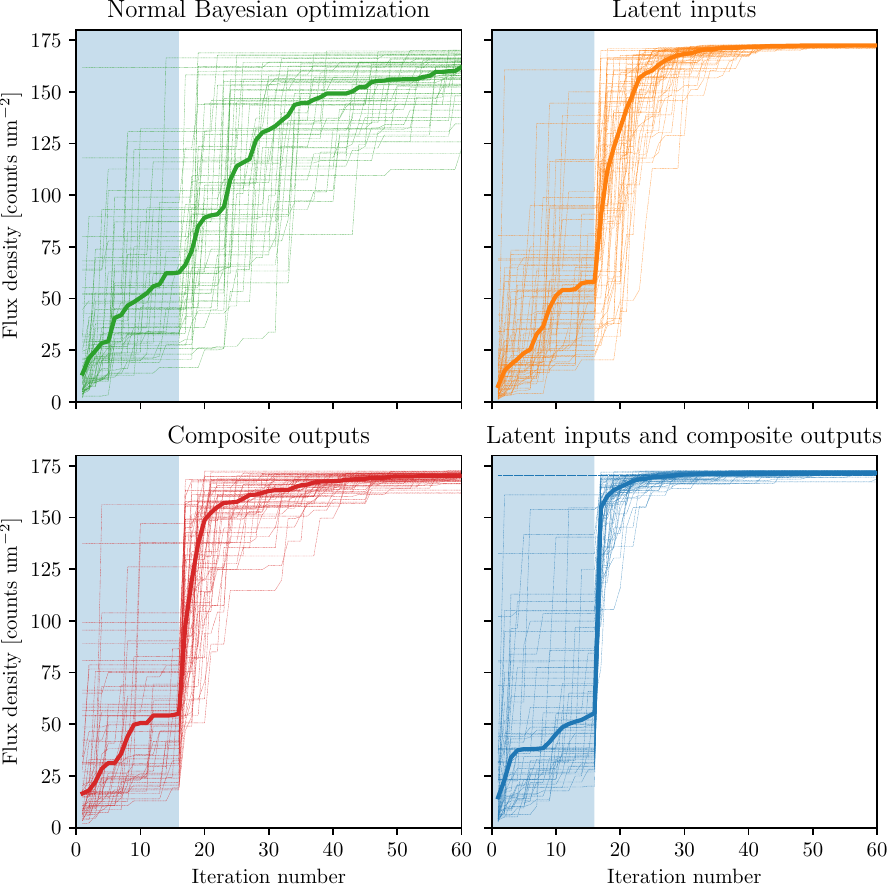}
\caption{The 4-dimensional optimization of just the K-B mirrors, whose motors are each misaligned by up 0.05\,mm. After an initial quasi-random sample of 16 points (shaded light blue), the agent is able to almost instantly return to the optimal alignment.}
\label{fig:tes-convergence-4d}
\end{figure}

The use of most beamlines is extremely competitive, and benchmarking alignment methods by performing ensembles of different runs is too time-intensive to be viable. Instead, we use digital twins of beamlines using the Sirepo-Bluesky backend \citep{rakitin2023recent}, allowing us to optimize the beam with the same Bluesky-based code used to align real beamlines. We use a ray tracing-based beamline simulation code called Shadow \citep{sanchez2011shadow3} to model beam propagation, which does not recreate diffraction effects but accurately recreates the behavior of the beam under misalignments. Even this heuristic method is slow, requiring several seconds per scan and thus many hours for comprehensive benchmarking; we note the development accelerated approximate models of beam propagation under misalignments, which would aid the efficient development of automated alignment tools \citep{nash2023reduced}.

For benchmarking, we consider the digital twin of the TES beamline at NSLS-II. In the 8-dimensional case, we use the six degrees of freedom outlined in Section~\ref{sec:real-tes}, but also allow the toroidal mirror to yaw and translate horizontally for a total of eight degrees of freedom. Each K-B motor can move up to $\pm$ 0.25\,mm from a fiducial starting point, while the range of each toroidal motor was bounded by the points where the misalignment of that motor caused the flux through the SSA to fall to 50\% of the maximum. The results of this benchmark are shown in Figure~\ref{fig:tes-convergence-8d}.

A simpler benchmark is shown in Figure~\ref{fig:tes-convergence-4d}, where the agent realigns the 4-dimensional K-B system under small misalignments (up to 0.05\,mm) in each mirror's motors.

\section{Further development and discussion}
\label{sec:discussion}

We have applied the same automated alignment tools to several different facilities, and have shown that the same python package can effectively align a range of beamlines.
Further refinement of these automated alignment tools will involve applying them to more beamlines at more facilities, with different flavors of optimization problems. How practical automated alignment can be necessitates an intuitive graphical user interface, from which the configuration of the optimizer is easy to understand. Further development also includes the implementation of new features and better performance in the software. The enabling of Pareto efficient optimization will give the beamline scientist more control over the beam quality, and making the agent take into account the traveling cost of moving the inputs into the acquisition function would allow for more informed optimization. We also plan to allow for a decentralized agent, which can run on a high-performance computing server and communicate with the control system using a streaming system like Kafka and feedback to the experiment control using Bluesky-Queueserver.\footnote{See \url{https://blueskyproject.io/bluesky-queueserver}}

Fly-scanning, the strategy of sampling while moving parameters (instead of stopping and settling at each input), presents the potential to speed up beamline alignment, as the sampling expense at most beamlines comes from the accelerating and decelerating of components while varying parameters. This requires a very accurate synchronization between the feedback of inputs and outputs (another use of the motor encoders mentioned in Section~\ref{sec:hysteresis}), and is actively being developed at many light source facilities. 

We also note that the largest obstacle to applying automated alignment to existing beamlines is the difficulty in constructing robust feedbacks, as many beam diagnostics have non-negligible backgrounds or malfunctioning pixels. While an experienced beamline scientist is able to ignore and look past these artifacts, they may interfere with simpler methods of estimating beam flux, position and size from an image (e.g. computing the spread of a profile summed along one dimension). This is especially significant in the case of Bayesian optimization, which relies on accurate sampling of the true objective. This suggests the benefit of more sophisticated diagnostic methods, using machine learning techniques like image segmentation.

\section{Acknowledgments}

The work was supported in part by BNL’s LDRD-22-031 project titled ``Simulation‐aided Instrument Optimization using Artificial Intelligence and Machine Learning Methods" and the DOE SBIR project (Award No. DE-SC00020593) titled ``X-ray Beamline Control with an Online Model for Automated Tuning and Reconfiguration".  
This research used 8-BM (TES) and 8-ID (ISS) beamlines of the National Synchrotron Light Source II, a U.S. Department of Energy (DOE) Office of Science User Facility operated for the DOE Office of Science by Brookhaven National Laboratory under Contract No. DE-SC0012704, and beamline 5.3.1 at the Advanced Light
Source, supported by the Director, Office of Science, Office of Basic Energy Sciences, of the U.S. Department
of Energy under Contract No. DE-AC02-05CH11231. 
AIW was partially supported by an Early Career Award in the X-Ray Instrumentation Program, in the Science User Facility Division of the Office of Basic Energy Sciences of the U.S. Department of Energy, under Contract No. DE-AC02-05CH11231.


\bibliography{refs.bib}

\end{document}